\documentclass[letter,twocolumn]{jpsj2} %% two-column layout
\def\runauthor{Hirono {\sc Kaneyasu} and Kosaku {\sc Yamada}}

\title{Theoretical Study on Coexistence of Ferromagnetism and Superconductivity}

\author{Hirono \textsc{Kaneyasu}$^{1}$\thanks{E-mail address:
hirono@sci.u-hyogo.ac.jp} and Kosaku \textsc{Yamada}$^{2}$}

\inst{$^{1}$Department of Material Science, University of Hyogo, Kamigori, Akou, Hyogo 678-1297, Japan \\
$^{2}$Department of Physics, Kyoto University, Kyoto 606-8502, Japan
}

\abst{On the basis of a two-dimensional $t$-$t'$ Hubbard model in ferromagnetic and paramagnetic states,
the triplet superconducting mechanism is investigated
by the third-order perturbation theory with respect to the on-site Coulomb interaction $U$. 
In general, the superconducting state is more stable in the paramagnetic state than in the ferromagnetic state.
As a special case, the dominant ferromagnetic superconductivity is obtained by
 the electron-electron correlation between the electronlike majority and holelike minority bands. 
Furthermore, it is pointed out that in some cases the two bands play an essential role 
for the coexistence of superconductivity and ferromagnetism.}

\kword{superconductivity, ferromagnetism, Hubbard model, UGe$_2$}

\begin{document}
\maketitle

%\section{Introduction} %% No sections necessary for express letters,
%letters and short notes

The experimental discoveries of uranium compounds have 
 attracted interest in the relation between ferromagnetism (FM) and
superconductivity (SC).
Generally, SC does not favorably coexist with FM since the FM moment gives rise to
 an internal magnetic field, which breaks the pairing state.
However, SC favorably coexists with FM in UGe$_2$ \cite{Saxena,Huxley} and
URhGe \cite{Aoki}. SC in these two materials is absent in the 
paramagnetic (PM) phase.
 Such a coexistence is rarely found in a few materials belonging to a
 strongly correlated system. 
Therefore, the coexistence may be realized only under
 certain specific conditions in the strongly correlated system.
To understand optimum conditions for the coexistence,
we investigate certain microscopic mechanisms between the 
coexistence and incompatibility in this paper. 
Here, that of SC in the FM state is similar 
to the situation of SC in the magnetic field. 
Arita $et$ $al$.\cite{Arita}
 studied the triplet SC induced in the magnetic field 
on the basis of the fluctuation exchange approximation. 
Their study is related to this investigation, although the methods and 
lattice structures are different. 

We focus our study on the possibility of the triplet 
pairing originating from the electron-electron correlation. 
The behavior of SC in PM and FM is studied in a quasi-
two-dimensional (2D) system. For this purpose, we adopt the 2D $t$-$t'$
Hubbard model with the majority and minority bands in the FM
state. 
The on-site Coulomb correlation $U$ works between the majority and
minority bands. 
The effective interaction for the triplet pairing is studied on the
basis of the perturbation theory with respect to the on-site Coulomb
repulsion $U$. The triplet SC in PM 
was analyzed by the third-order perturbation theory (TOPT)\cite{Nomura}.
 We extend the theory to SC in the FM state. 
The dependence of SC on the
magnetization is studied using the Eliashberg equation in the FM state. 

In this model, SC is not induced in the complete
FM state corresponding to the empty in the minority bands, 
since the effective interaction vanishes between the two bands. 
With decreasing FM moment, the effective interaction
has the momentum dependence due to $U$ between the majority and minority 
bands. The momentum dependence is sensitive to changes in the two
bands.
The value of magnetization corresponds to the difference between
filling numbers in the two bands. 
Thus, the change in magnetization leads to the changes in 
both Fermi surfaces (FS) and the density of states (DOS) in the two bands.
In the above consideration, we study the relation between SC and the change of bands from PM to FM.

%\section{Formulation}
The formulation is as follows. 
The 2D $t$-$t'$ Hubbard Hamiltonian with the two bands is given by
\begin{equation}
{\cal H}=t_{s} \sum_{{\bf i},{\bf a},s} c^{\dag}_{{\bf
i},s}c_{{\bf i+a},s}+t'_{s} \sum_{{\bf i},{\bf b},s}
c^{\dag}_{{\bf i},s}c_{{\bf i+b},s}+U \sum_{{\bf i}}n_{{\bf
i}, s} n_{{\bf i}, -s},
\end{equation}
where $c_{{\bf i},s}$ is an annihilation operator for a quasi
particle with spin $s$ at site ${\bf i}$. ${\bf a}$ and ${\bf b}$
are, respectively, the vectors connecting nearest-neighbor and
next-nearest-neighbor sites in a square lattice. 
The indices $s$ with 
$\uparrow$ and $\downarrow$ indicate majority
and minority bands, respectively.
The transfer integrals $t_{s}$ and $t'_{s}$ denote nearest-neighbor and
next-nearest-neighbor hopping integrals, respectively. $U$ is the
on-site Coulomb interaction between majority and minority electrons, and $n_{{\bf i},s}$=$c^{\dagger}_{{\bf i},s}c_{{\bf i},s}$.
We assume the dispersion $E_{\bf \it k,s}$ so as to produce the
 2D cylindrical majority FS and minority FS in the square lattice,
leading to 
\begin{equation}
E_{\bf \it k,s}=2t_{s}(\cos k_x+\cos k_y)+4t'_{s}\cos k_x \cos k_y. 
\end{equation}
Then, we obtain the bare Green's function as
$G_{0,s}(k)$=$1/[i\omega_n-(E_{k,s}-\mu_{0,s})]$, where
$k$ is the short-hand notation defined as $k$=$({\bf k},\omega_n)$,
${\bf k}$ is the momentum and $\omega_n=\pi T(2n+1)$ is the fermion
Matsubara frequency with temperature $T$. Note that the chemical
potential $\mu_{0,s}$ for the noninteraction case is determined by the
electron number $n_{s}$ (per site and spin) as $n_{s}$=$\sum_{k}G_{0,s}(k)$, where $\sum_k$=$(T/N)\sum_{\bf k}\sum_n$ and $N$ is the number of sites. 
The difference between filling numbers in majority and minority bands
corresponds to the value of magnetization. The filling numbers are changed from
PM to FM by the shifts of the chemical potential $\mu_{0,\uparrow}$
and $\mu_{0,\downarrow}$ with the constant value of the total
electron number $n_{\uparrow}+n_{\downarrow}$. 
\begin{figure}[tb]
\begin{center}
\includegraphics[height=90mm]{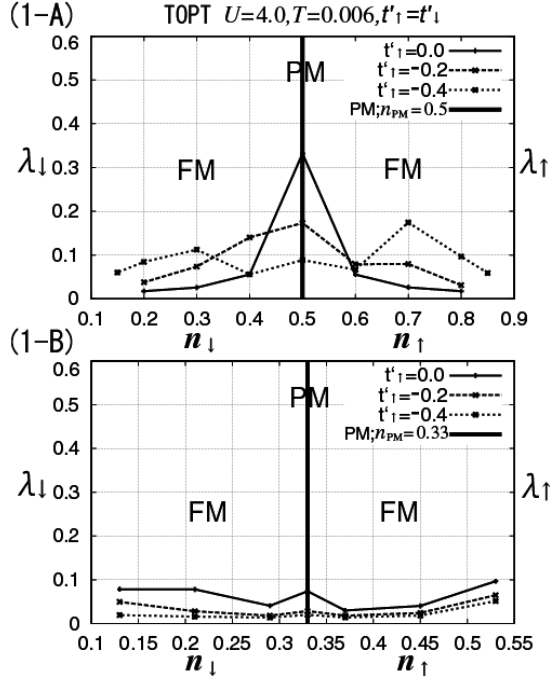}
\end{center}
\caption{Dependence of $\lambda$ on $n_{\uparrow}$ 
and $n_{\downarrow}$ in case (1). 
$n_{\rm PM}=n_{\uparrow}=n_{\downarrow}$ in PM equals 
the half-filling and 0.33 far from the half-filling, corresponding to
(1-A) and (1-B), respectively. $\lambda$ values of majority and
minority bands correspond to the right- and left-hand side of the figure, respectively.}
\label{f1}
\end{figure}
\begin{figure}[tb]
\begin{center}
\includegraphics[height=90mm]{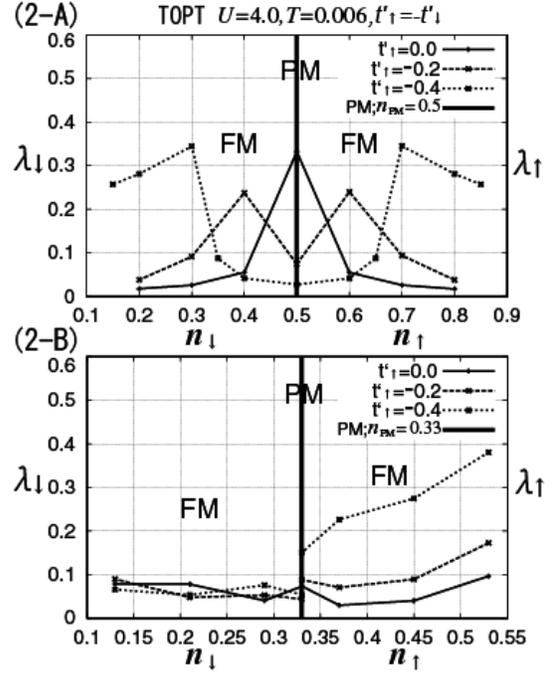}
\end{center}
\caption{Dependence of $\lambda$ on $n_{\uparrow}$ 
and $n_{\downarrow}$ in case (2). $n_{\rm PM}$ 
in PM equals the half-filling and 0.33, 
corresponding to (2-A) and (2-B), respectively.}
\label{f2}
\end{figure}

For the triplet pairing, the effective interactions $V$ for the
two bands are based on the TOPT extended from PM to FM, 
and they are given by
\begin{equation*}
\hspace{-3.0cm}{\it V}^{\rm TOPT}_{s}(k,k')=U^2\chi_{0,-s,-s}(k-k')
\end{equation*}
\vspace{-0.6cm}
\begin{equation*}
\hspace{1.0cm}+2 U^3 Re \sum_{k''}G_{0,-s}(k+k''-k')(\chi_{0,s,-s}(k+k'')
\end{equation*}
\vspace{-0.6cm}
\begin{equation}
\hspace{4.2cm}+\phi_{0,s,-s}(k+k''))G_{0,-s}(k''),
\end{equation}
with
\vspace{-0.3cm}
\begin{equation}
\chi_{0,s,s}(q)=-\sum_{k}G_{0,s}(k)G_{0,s}(q+k),
\end{equation}
\vspace{-0.3cm}
\begin{equation}
\phi_{0,s,s}(q)=-\sum_{k}G_{0,s}(k)G_{0,s}(q-k).
\end{equation} 
An effective pairing interaction $V^{\rm TOPT}_{s}$ between particles
is evaluated using TOPT. To clarify in detail the effect of the perturbation term, 
we also study SC on the basis of the second-order perturbation theory
(SOPT).
An anomalous self-energy $\Sigma_{A,s}$ is expressed using $V_{s}(k,k')$
and an anomalous Green's function $F(k)$ as 
$\Sigma_{A,s}(k)$=$-\Sigma_{k'}V_{s}(k,k')F_{s}(k')$. At the SC transition temperature $T_{\rm SC}$, the
linearized Eliashberg equation is reduced to the eigenvalue equation, 
\vspace{-0.1cm}
\begin{equation}
\lambda_{s}\Sigma_{A, s}^{\dagger}(k)=-\sum_{k'}V_{s}(k,k')|G_{s}(k')|^2\Sigma_{A, s}^{\dagger}(k').
\end{equation}

We take $|t_{s}|$=1.0 as the unit of energy. 
The large $\lambda$ value indicates that $T_{\rm SC}$ becomes high. 
An effective pairing interaction $V_{s}$ is also evaluated on the
basis of TOPT. Although the origin of SC is investigated using the
total terms in $V_{s}$, in order to analyze the role of $V_{s}$ in detail, it
is convenient to divide it into two parts, namely, $V_{s}^{TOPT}$ and $V_{s}^{SOPT}$.; 
$V^{\rm SOPT}_{s}$ and $V^{\rm TOPT}_{s}$ represent the effective
interaction obtained on the basis of SOPT and TOPT, respectively. 
We solve the equation with the assumption that
$\Sigma_{A, s}^{\dagger}$ has a triplet symmetry.
In the numerical calculation, we divide the first Brillouin zone (BZ) into 128$\times $128
momentum meshes and take $N_{f}$ = 1024 for Matsubara frequency
$\omega_n$. The bandwidth $W$ ($W\sim 8 t$) is a necessary
range of $\omega_n$ for reliable calculations. The range is covered
under the condition $|W|< \pi T N_{f}$. To satisfy this condition,
we calculate $\lambda$ in the region with $T>$0.004.
%\section{Results}

First, the behavior of $\lambda$ is investigated for the following
case, where FS ((1),(2)) and the filling number ((A),(B)) in PM and FM
are assumed as follows.
\\ 
(1) Electronlike majority FS and electronlike minority FS 
(Transfer integrals $t_{s}$ and $t'_{s}$ have the same signs between the majority and minority bands.)

(2) Electronlike majority FS and holelike minority FS 
(Transfer integrals have the opposite signs between the two bands.)

Here, we take $t_{\uparrow}=-1.0$ and we choose the absolute values of
$t_{s}$ and $t'_{s}$ to be identical between the two bands for simplification.

(A) Half-filling in PM 
(Filling number in PM: $n_{\rm PM}=(n_{\uparrow}+n_{\downarrow})/2=n_{\uparrow}=n_{\downarrow}$=0.5) 
\\
(B) Filling number far from half-filling in PM 
($n_{\rm PM}$=0.33)

The filling numbers of the two bands are changed from PM to FM,
starting from the filling number in PM. Here, PM indicates that the filling number agrees between the majority and minority bands. PM in case (1) is the general PM state with the same FS in the two bands. 
On the other hand, PM in case (2) is not the general PM and possesses a different FS for the two bands. 
\begin{figure}[tb]
\begin{center}
\includegraphics[height=105mm]{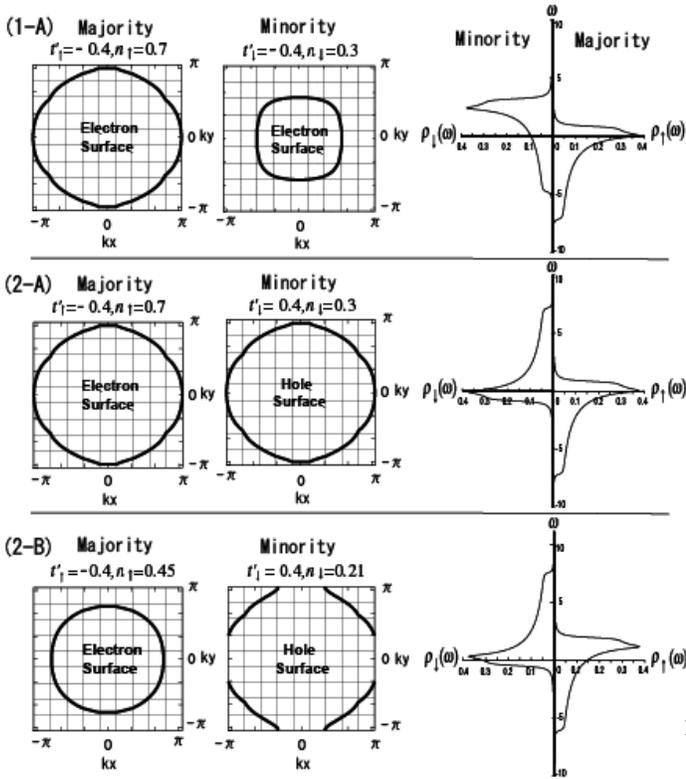}
\end{center}
\caption{FS in BZ and DOS.}
\label{f3}
\end{figure}

As the results, for cases (1-A), (1-B), (2-A) and (2-B), the behavior
of $\lambda$ obtained on the basis of TOPT is shown in Figs. 1 and 2.
$\lambda$ in PM has identical values between (1-A) and (2-A) due to the agreement of FS in the two bands. 
The large $\lambda$ for (1-A) is given in PM, as shown in the data for $t'_{\uparrow}=0$.
(1-B) has a small $\lambda$ value in both PM and FM, thus the case is inappropriate for the realization of SC. 
$\lambda$ for (2-A) takes a large value at $t'_{\uparrow}=-0.4$ in FM.
(2-B) takes the largest value of $\lambda_{\uparrow}$ among the four cases. 
Thus, (2-B) gives the highest $T_{\rm SC}$ in FM among the four
cases. The tendency just exists at other two $t'_{s}$ values. 
From these results, we can say that SC is more favored in the FM state
than in the PM state for the combination of electronlike FS and holelike FS
 (Case (2)), while that is not the case for the combination of
electronlike FS and electronlike FS (Case (1)). 
To add to the appropriate combination of the electronlike and holelike FS, the possibility of SC coexisting with FM is enhanced 
by the filling number far from the half-filling in PM (Case (2-B)).
We give a comment about the dominant band for SC, which gives a large
contribution to the increase in $\lambda$.
The majority band is dominant in SC in Figs. 1 and 2. However, the minority
band is dominant in SC in the case of $n_{\rm PM}$=0.67. The case relates to that of $n_{\rm PM}$=0.33 as the electron-hole symmetry. 
Thus, the determination of the dominant band depends on the features of the two bands. 
Therefore, the dominant band in SC is not always the majority band. 

Next, we detail the effect of the two bands.
In the Eliashberg equation, the $\lambda$ value is sensitive to
$G_{\uparrow}G_{\downarrow}$ near the Fermi level. 
$G_{\uparrow}G_{\downarrow}$ depends on both DOS and FS in the two bands.
Therefore, it is important to consider the effects of FS and DOS on the effective interactions ($V^{\rm TOPT}$ and $V^{\rm SOPT}$) 
including $G_{\uparrow}G_{\downarrow}$. We investigate the relations between these details and the behavior of $\lambda$.
$\lambda_{\uparrow}$ for (1-A) has a large value at $t'_{\uparrow}=-0.4$ in FM in Fig. 1.
DOS $\rho_{s}$ in Fig. 3 is related to FM-SC.
The large $\lambda$ value is mainly contributed by the large DOS value
at Fermi level in the majority band, as shown in Fig. 3(1-A).
Corresponding to this, at the same $|t'_{s}|$ value in case (2-A), 
both the majority and minority bands have a large DOS value at the Fermi level, as shown in Fig. 3(2-A).
Therefore, the $\lambda$ value in case (2-A) is larger than that in
case (1-A). 
In Fig. 2(2-B), the large $\lambda_{\uparrow}$ value at $t'_{\uparrow}=-0.4$ in FM near $n_{\uparrow}=$0.45
is mainly contributed by the large DOS value at the Fermi level in the minority band, as shown in Fig. 3(2-B).
The large DOS value is due to the location of the van Hove
singularity at the Fermi level. 
The large DOS value at the Fermi level leads to a strong electron
correlation, 
which is the origin of SC in FM. Thus, the large DOS value at
the Fermi surface is also essential to 
 the mechanism of FM-SC, as well as that of PM-SC.
\begin{figure}[tb]
\begin{center}
\includegraphics[height=90mm]{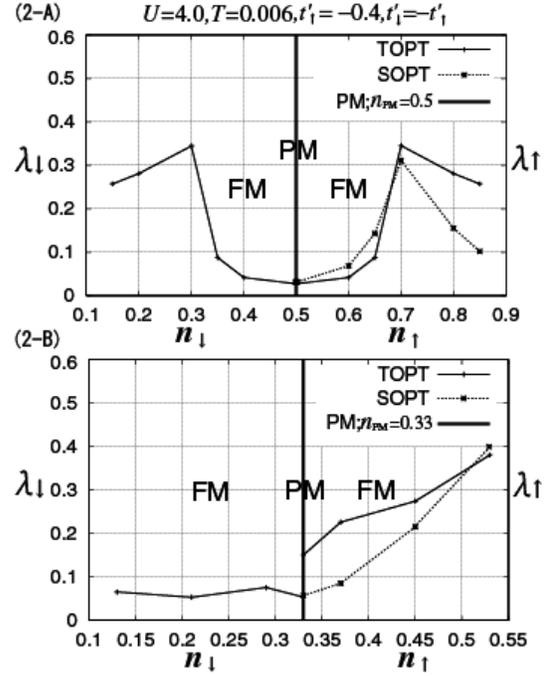}
\end{center}
\caption{Dependence of $\lambda$ on $n_{\uparrow}$ and
$n_{\downarrow}$, which is obtained on the basis of TOPT and SOPT.}
\label{f5}
\end{figure}
\begin{figure}[tb]
\begin{center}
\includegraphics[height=90mm]{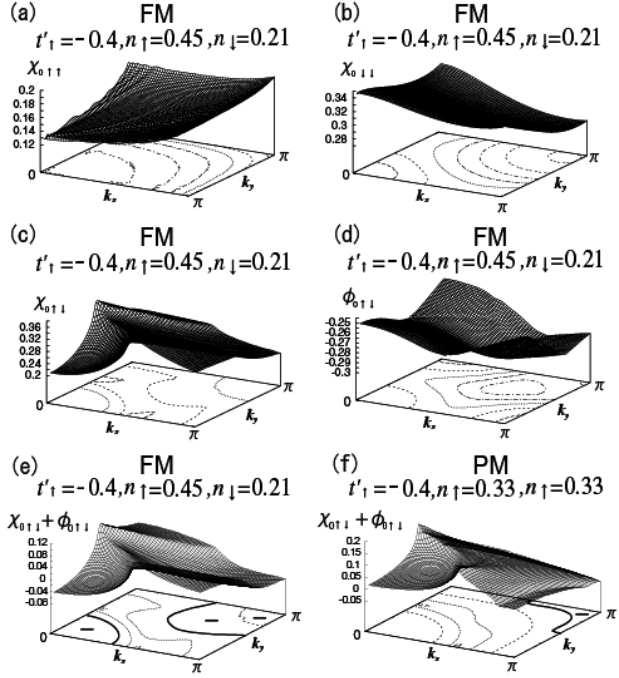}
\end{center}
\caption{$\chi_0(q, \omega_n=0)$, $\phi_0(q, \omega_n=0)$ and $\chi_0(q, \omega_n=0)+\phi_0(q, \omega_n=0)$ for (2-B).}
\label{f6}
\end{figure}

Next, we explain the effect of the momentum dependence in the
effective interaction, which originates from the features of FS. 
The FS features are shown in Fig. 3.
In the combination of electronlike FS and electronlike FS, which does not
lead to the larger value of $\lambda$ in FM than in PM, 
the momentum dependence suitable for the triplet pairing is induced by
the identical FS features of the two bands in PM. However, the discrepancy of the two bands in FM does not induce the momentum dependence suitable to the triplet pairing.
On the other hand, in the combination of electronlike FS and holelike FS, 
 the large $\lambda$ value is obtained in FM. 
To clarify the effect of the momentum dependence, 
the comparison between TOPT and SOPT for (2-B) is shown in Fig. 4.
Furthermore, the momentum dependences of $\chi_0$ and $\phi_0$ for (2-B) 
are shown in Fig. 5. 
For (2-A) in Fig. 4, $\lambda_{\uparrow}$ of SOPT at $t'_{\uparrow}=-0.4$ has a large value in FM. 
The behavior accords with the features of DOS.
In contrast to this, for case (2-B) shown in Fig. 2, the influence of DOS is ambiguous
in the strong FM state ($n_{\uparrow}>$0.45) at $t'_{\uparrow}=-0.4$. $\lambda_{\uparrow}$ for SOPT at $t'_{\uparrow}=-0.4$
 becomes large as the magnetic moment becomes large. $\lambda$ for
 TOPT takes a higher value than that for SOPT 
with the filling number far from the half-filling. These behaviors do
not accord with the large DOS value at the Fermi level.
This finding indicates that the momentum dependence originating from FS is important for FM-SC.
$V_{\uparrow}^{SOPT}$ is contributed by the momentum dependence 
of $\chi_{\downarrow, \downarrow}$ with the hump around $(k_x, k_y)=(0,0)$, as shown in Fig. 5(b). The momentum dependence includes 
the FM spin fluctuation. The spin fluctuation is also included in the complex momentum dependence of $\chi_{\uparrow, \downarrow}$
, which is reflected in the third-order terms. 
The term including $V^{\rm TOPT}$ is also contributed by $\chi_{0,s,-s}+\phi_{0,s,-s}$. 
Thus, the two bands are reflected in the third-order term. 
$\phi_{0,s,-s}$ in FM with the different two FS features has a more complex momentum dependence than 
that in PM, which indicates the complete agreement of the two FS in case (1). 
 The region of the negative sign in the momentum dependence of $\chi_{0,s,-s}+\phi_{0,s,-s}$
has dominant effects on SC in the third-order term. 
Case (e) for FM has a larger $\lambda$ value than case (f) for PM
 because case (e) has a larger part with the negative sign than case (f).
In the SOPT and TOPT, the momentum dependence of
 $V_{\rm SOPT}$ and $V_{\rm TOPT}$ originating from FS determines FM-SC.

Here, we discuss the generality over wide regions of the parameters. 
The effects of DOS and momentum dependence of the majority and minority bands on FM-SC are also dominant on FM-SC 
in the other filling numbers and hopping integrals. 
The importance of the effects is general over wide regions of the parameters. 
In this paper, we show the representative behaviors at the parameters of the four cases.
Finally, we mention the pairing symmetry obtained in $\Sigma_{A, \uparrow}$. 
In the parameter with the largest $\lambda$ value for (2-B), the momentum
dependence of the pairing symmetry indicates the $p_x$-wave symmetry in FM. 
 
%\section{Summary}
As the summary, we conclude that the mechanism of FM-SC depends on the 
features of both majority and minority bands. 
 (I) The band with the largest DOS at the Fermi level gives the dominant effect on the rise of $T_{\rm SC}$ in FM. 
The location of the large DOS at the Fermi surface gives rise to the strong correlation in FM.
 (II)  The momentum dependence of FS of the two bands in FM is important for 
the momentum dependence of the effective interaction to induce the attractive interaction for SC. 
(III) The existence of both majority and minority spins is essential for the origin of FM-SC. 
FM-SC is not induced in the absence of the minority spins corresponding to the full magnetized state,
 because of the absence of the electron-electron correlation between
the majority and minority bands. 
(IV) SC is more favorable in FM rather than in PM in the combination
of electronlike majority FS and holelike minority FS. 
The possibility of SC coexisting with FM is strong in a specific
situation with the combination of electronlike FS and holelike FS, 
which have the filling number far from half-filling in PM. 
On the other hand, the realization of FM-SC is difficult
 in the combination of electronlike FS and electronlike FS or under the condition near half-filling in PM.
These findings indicate that the coexistence of FM and SC happens due to the specific features of the two bands. 
Therefore, these findings explain that the coexistence of FM and SC is a rare phenomenon. 

Here, we comment on FM-SC in UGe$_2$ in the strongly correlated systems.
Recently, the band calculation in PM indicates the two bands with the filling number far from the half filling.\cite{Biasini,Yamagami,Shick} 
The effective combination of the two bands is the electronlike FS and holelike FS, which possess the almost 2D features in PM. 
We assume the following situation in FM.
In the FM phase, the electronlike FS in PM may split into the majority
electronlike FS and minority electronlike FS.
As well as this, the holelike FS in PM may split into the majority
holelike FS and minority holelike FS.
Under this assumption, we should compare the eigenvalues between the four bands in FM and the two bands in PM.
The majority FS and minority FS in PM are not identical in case
(2), and the results in (2) is insufficient for SC in PM. 
In adding to this, both center points of the two FS features in PM locate at the
center of BZ of our model, and these are different from those located near
the center and edge of BZ obtained by band calculation. Therefore, the
result may change in the model with different center points. This
problem is our further investigation.\cite{Zigzag} However, in the region of FM states, the result of (IV)
may suggest the following explanation.
The triplet SC in FM is mainly induced by
the electron correlation between the majority electronlike (holelike) and minority holelike (electronlike) bands 
rather than by that between the majority electronlike (holelike) and minority electronlike (holelike) bands. 
Since we have holelike and electronlike Fermi
surfaces, we can expect that the mechanism (IV) is effective. 
Experimentally, the markable mass enhancement is observed
 in the region of the pressure-induced SC, 
and the mass enhancement indicates the effect of the strong correlation for the origin of SC in FM.\cite{Yamagami}

From our results in this simple model, the electron correlation is expected to be strong 
by the large DOS value at the Fermi level, which is obtained by the markable change of the magnetization, 
in addition to the momentum dependence including the FM spin fluctuation.
Therefore, as in the system, the change of FS of the two bands in FM
is related to the mechanism of FM-SC. 
The existence of the minority spin is essential for the appearance of FM-SC. We suggest the possibility of the $p$-wave
triplet FM-SC in UGe$_2$ on the basis of the electron-electron correlation 
between the majority and minority spins.

%\section*{Acknowledgment}
The authors thank Professor. T. Yamagami, Professor. K. Makoshi and Dr. H. Ikeda for valuable discussions.


\begin{thebibliography}{99}
\bibitem{Zigzag} We had investigated 
the problem in the structure of zigzag-chain. The result supports
 the conclusions (I)-(IV) of our this paper.
\end{thebibliography}
\end{document}